**Band-like Electron Transport with Record-High Mobility in the TCNQ family**

*Yulia Krupskaya[1]\*, Marco Gibertini[2]\*, Nicola Marzari[2]\*, and Alberto F. Morpurgo[1]\**

[1]Department of Quantum Matter Physics (DQMP), University of Geneva, 24 quai Ernest-Ansermet, CH - 1211 Geneva, Switzerland
[2]Theory and Simulation of Materials (THEOS) and National Center for Computational Design and Discovery of Novel Materials (MARVEL), École Polytechnique Fédérale de Lausanne, MXC 340 (Bâtiment MXC), Station 12, CH-1015 Lausanne, Switzerland

E-mail:
Dr. Y. Krupskaya: y.krupskaya@ifw-dresden.de
Prof. A.F. Morpurgo: alberto.morpurgo@unige.ch
Dr. M. Gibertini: marco.gibertini@epfl.ch
Prof. N. Marzari: nicola.marzari@epfl.ch

It is believed that in the absence of extrinsic disorder, electronic transport in field-effect transistors (FETs) of top-quality organic molecular semiconductors occurs in the so-called band-like regime, in which the carrier mobility increases upon lowering temperature.[1-9] To understand the nature of this regime and to identify the microscopic material parameters that favor its occurrence –both currently unknown–[10] comparative studies of different organic semiconductors are needed. It would be ideal to compare materials formed by identical constituent molecules, but having different crystal packing, resulting in different electronic bands, molecular motion, and coupling between the two. In principle, polymorphs of a same molecule (for instance pentacene)[11, 12] could be used to this end, but in practice polymorphism is more commonly found in thin-films of organic semiconductors,[11, 12] whose quality is not sufficient for the experimental investigation of band-like transport.[13] Another possibility is to identify families of molecules that –although not identical– are sufficiently similar to allow exploring correlations between the transport properties observed experimentally and different microscopic aspects of their crystalline and electronic structure. A family of molecules suitable for this study has not yet been identified, largely because the number of organic molecular semiconductors known to exhibit band-like transport is limited. Here, we show experimentally that the $F_x$-TCNQ (fluorinated tetracyanoquinodimethane) molecules do constitute a suitable family, and compare the transport properties of TCNQ, $F_2$-TCNQ, and $F_4$-TCNQ single-crystals using field-effect transistors.

Our work is enabled by a main, unanticipated finding. $F_2$-TCNQ single-crystal devices reproducibly exhibit very high room-temperature electron mobility $\mu$ = 6-7 cm$^2$V$^{-1}$s$^{-1}$, increasing by a factor of 3 to 4 upon reducing $T$ from room temperature to 150 K, where it reaches a value $\mu$ = 25 cm$^2$V$^{-1}$s$^{-1}$ (the largest relative increase in mobility upon lowering $T$ ever reported in organic FETs). This finding is unexpected because TCNQ single-crystal FETs had been investigated earlier, and found to be of insufficient quality to access the band-like transport regime;[14, 15] similarly, our measurements on $F_4$-TCNQ devices exhibit no sign of band-like transport. As a first step to analyze the properties of the three different materials, and to understand why $F_2$-TCNQ FETs exhibit such an outstanding behavior, we have performed systematic structural studies and electronic band-structure calculations. We find that $F_2$-TCNQ crystals have a crystalline structure with a primitive unit cell containing a single molecule (very unusual for organic molecular crystals),[16] so that all the molecules in the crystal are oriented parallel to each other. Additionally, the band originating from the overlap between LUMOs (lowest unoccupied molecular orbitals) has a very large width (the largest among the organic molecular semiconductors commonly used for the realization of FETs)[16, 17] and a clear three-dimensional character. These properties are conducive to a more pronounced electron delocalization,[2, 9, 18] which could account for the remarkable behavior of $F_2$-TCNQ single-crystals FETs.

Electronic transport measurements were performed on $F_2$-TCNQ single crystals in a vacuum gap FET configuration.[14] **Figure 1**a shows an optical-microscope image of a typical device, in which a polydimethylsiloxane (PDMS) stamp is used to suspend a single crystal on top of a recessed gate electrode. The shape of the crystals is needle-like with an unusual, approximately square cross-section, enabling transport measurements to be performed on two different crystal facets (i.e., FETs can be assembled so that either one or the other surface faces the gate electrode). The electrical device characterization, performed by means of conventional transistor measurements, is shown in Figure 1b – 1d. It is apparent from the evolution of the slope in the FET transfer curves (i.e., the source-drain current $I_{SD}$ measured as a function of gate voltage $V_G$, at a fixed source-drain bias $V_{SD}$; Figure 1b) that the carrier mobility increases upon lowering $T$ between room temperature and approximately 150 K, while it starts decreasing below this temperature. The output ($I_{SD}$ vs. $V_{SD}$ at fixed $V_G$) and transfer curves for the same device, measured at the temperature of maximum mobility (154 K), are presented in Figure 1c and 1d. They show virtually ideal behavior (absence of hysteresis, no sign of contact effects), except for the presence of background bulk conductance due to unintentional doping, which is also clearly visible at negative gate voltages in the transfer curves (Figure 1b) and is responsible for the incomplete saturation of the output curves (Figure 1c). The amount of unintentional doping –roughly estimated from the background conductance measured at large negative $V_G$ values– was found to vary from device to device, without any correlation to the mobility value (a large background conductance was also observed in TCNQ and $F_4$-TCNQ devices).

Figure 1e shows the behavior of the electron mobility extracted for three different $F_2$-TCNQ transistors from the derivative of the source-drain current with respect to the gate voltage $dI_{SD}/dV_G$. In all cases the mobility $\mu$ (6-7 $cm^2V^{-1}s^{-1}$ at room temperature) increases upon cooling (we checked that measurements done upon cooling and warming up give identical results, with no hysteresis); it peaks at $T \sim 150$ K, where it reaches values as high as 25 $cm^2V^{-1}s^{-1}$ in the best device and is always higher than 15 $cm^2V^{-1}s^{-1}$. These values represent a large improvement over the best existing $n$-type organic FETs[6, 8, 19, 20, 21, 22] and, at low temperature, they are less than a factor of two smaller than those measured in the highest-quality $p$-type Rubrene single-crystal transistors ever reported.[1] Interestingly, the experiments reveal that the same value of carrier mobility is measured on devices in which the FET channel is on either one of the two different crystalline surfaces, a very unusual finding for organic semiconductors.

The outstanding transport properties of $F_2$-TCNQ crystals are not matched by TCNQ and $F_4$-TCNQ single-crystal FETs. The output and transfer characteristics for TCNQ and $F_4$-TCNQ FET are presented in **Figure 2**a, 2b and Figure 2d, 2e, respectively. We find a carrier mobility of $\mu \sim 0.1$ cm$^2$V$^{-1}$s$^{-1}$ for TCNQ and $\mu \sim 0.2$ cm$^2$V$^{-1}$s$^{-1}$ for $F_4$-TCNQ crystals at room temperature, decreasing upon cooling (see Figure 2c and 2f). We have realized many different PDMS stamp FETs with both TCNQ and $F_4$-TCNQ single crystals, and found that the behavior of these devices is extremely reproducible, just as it is the case for $F_2$-TCNQ. We note that the mobility that we found for TCNQ devices is lower than that reported by Menard *et al.*[14], a discrepancy whose precise origin is not clear. The important point, however, is that in both the devices of Menard *et al.* and ours, the mobility of electrons in TCNQ crystals is found to be thermally activated.

There does not appear to be any obvious extrinsic reason to account for the observed large disparity in the transport properties of these three materials. It may be that the chemical purity of $F_2$-TCNQ is better than that of TCNQ and $F_4$-TCNQ, a possibility that is difficult to rule out conclusively, because it is virtually impossible to determine the concentration of remnant impurities in organic semiconductors. However, indirect experimental observations (the amount of residue observed during the crystal growth or the analysis of the off FET current, which provides an estimate of the doping level in the materials) suggest that the concentration of impurities is comparable in all crystals of the three molecules, and provide no evidence for the large difference observed in transport. We have also performed extensive atomic force (AFM) microscopy imaging to rule out that the different transport properties of the three materials originate from differences in the quality of the crystal surfaces (where conduction in FET devices occurs). Indeed, **Figure 3** shows that the surfaces of crystals of the three molecules all exhibit large, highly-ordered terraces, separated by steps of quantized height (the height –see Figure 3g, 3h, 3i for TCNQ, $F_2$-TCNQ, $F_4$-TCNQ, respectively– is in very good agreement with the crystal unit cell parameters obtained from the X-ray diffraction, see Table S2 in the supporting information).

In the absence of obvious differences in quality or purity of the different materials, we proceed to analyze their crystalline and electronic structure, to identify the origin of the outstanding transport properties of $F_2$-TCNQ. We performed X-ray crystal structure measurements for TCNQ, $F_2$-TCNQ and $F_4$-TCNQ crystals. For all three materials we observe a single phase without any sign of polymorphism. To analyze the possibility that a phase transition in $F_2$-TCNQ crystals occurs at $T \sim 150$ K, where we observe a sharp downturn of the mobility upon further cooling, we performed extensive measurements at $T = 100$ K and 180 K, to fully solve the crystal structure at both temperatures. We found no change in the crystal structure symmetry and no indications of a phase transition. This conclusion is further supported by the linear behavior of the unit cell volume vs. *T*, which we found from room temperature down to $T = 100$ K (see Figure S2 in the supplementary information). In presenting the results of the structural analysis, we focus on the molecular packing at the surface corresponding to the plane onto which electrons are accumulated in the FET devices (the transport plane, coincident with face 2, according to the notation of **Figure 4**a). To illustrate the crystal structure in the third dimension, we also show the molecular packing in the plane corresponding to the surface that is (approximately) perpendicular to face 2 (indicated with face 3 in Figure 4a). The experimentally determined molecular packing on face 2 and face 3 is shown for the three materials in Figure 4c – 4e, and Figure 4f – 4h, respectively.

Two important observations can be made. The first is that the molecular packing on face 2 and 3 of $F_2$-TCNQ crystals is identical, which immediately explains why $F_2$-TCNQ devices assembled using the two different crystal surfaces result in the same electron mobility values, as we stated earlier. The second is that there is only one molecule per primitive unit cell in $F_2$-TCNQ crystals, so that all molecules are parallel to each other, stacked in a face-to-face fashion. Both aspects are rather unique. Among all organic molecular semiconductors commonly employed in the realization of FETs, the molecular packing on the different crystal surfaces is extremely different, with high-mobility transport occurring only on one of them. Additionally, the unit cell is normally formed by two or more molecules.[16] Indeed, this is also the case of TCNQ and $F_4$-TCNQ, which have respectively two and four molecules in the primitive unit cell of their crystals and clearly different packing on their surfaces (see Figure 4c, 4f for TCNQ and Figure 4e, 4h for $F_4$-TCNQ).

Having determined the crystal structure, we perform first-principles calculations of the atomic and electronic structure of TCNQ, $F_2$-TCNQ, and $F_4$-TCNQ using density-functional theory with van-der-Waals corrected functionals.[23] As mentioned in the supplementary material, the structural parameters are in excellent agreement with experiments. Then, we looked for each material at the bands stemming from LUMO overlaps, i.e., the bands hosting the electrons accumulated in the FET channel (see supplementary information for details). The band dispersion along high-symmetry directions in the Brillouin zone is shown in **Figure 5**. We see that $F_2$-TCNQ crystals have an extremely large (for van der Waals bonded organic semiconductors) band-width $W_{F2}$ = 0.8 eV, larger than in any other organic molecular semiconductors employed in the realization of FET devices.[16, 17] As a term of comparison, in Rubrene crystals –the organic semiconductor exhibiting the largest hole mobility in *p*-type transistors– the width of the highest occupied molecular orbital in which holes are accumulated, is $W_{rub}$ ~ 0.4 eV.[24, 25]

From the analysis of the crystal structure and from the band-structure calculations it follows that $F_2$-TCNQ crystals obey all criteria that are "canonically" considered as favorable to achieve high-quality transport in organic semiconductors.[17] These include an unprecedented width of the band in which carriers are accumulated, a molecular packing resulting in a three-dimensional electronic coupling that is conducive to an enhanced delocalization (i.e., $F_2$-TCNQ crystals are not electronically quasi two-dimensional as virtually all other organic semiconductors),[16] and a face-to-face packing of all molecules, which may be argued to decrease the coupling of electronic and molecular motion. All these material properties are consistent with the experimentally observed outstanding transport in $F_2$-TCNQ FETs. What is more difficult to understand is why the two other materials do not perform better than they actually do. This is particularly the case for TCNQ, whose band-width, $W_{TCNQ}$ ~ 0.5 eV (see Figure 5a), is still large (even slightly larger than that of Rubrene), and whose molecular packing in the transport plane is very similar to that of $F_2$-TCNQ. Nevertheless, the room-temperature carrier mobility is approximately 50 times smaller than in $F_2$-TCNQ FETs –even slightly smaller than in $F_4$-TCNQ– and transport is thermally activated, not band-like. Therefore, even though the considerations based on the static molecular packing and electronic band structure do (correctly) indicate that $F_2$-TCNQ crystals are ideally suited to achieve top quality transport, they cannot rationalize the comparative behavior exhibited by the three materials and explain why, with crystals of comparable quality, $F_2$-TCNQ devices exhibit an unprecedented increase in mobility[1-8] upon cooling, whereas in TCNQ devices transport is thermally activated (i.e., transport is dominated by extrinsic effects).

In this respect, the different couplings of the charge carriers (electrons in the present case) in the three materials with the vibrational degrees of freedom can play a dominant role. Indeed, it is well known[16] that electron-phonon interactions are especially relevant for organic crystals, where weak van-der-Waals interactions give rise to many low-energy vibrational modes that are thermally excited in the temperature range of the experiments. In light of the present results, we note that in TCNQ, $F_4$-TCNQ, and in the majority of other organic crystals the unit cell contains more than one molecule. While this does not necessarily affect the phonon frequencies (in all cases these organic crystals have many phonon bands below 200 cm$^{-1}$, all excited at room temperature), it does create a more diverse set of couplings, making it more likely for the electron or hole pockets to be strongly affected by some of these different vibrational degrees of freedom. A stronger coupling effectively leads to a larger "dynamical disorder" induced by the thermally excited molecular motion,[2, 9, 18] which is experienced as static by the charge carriers, and results in a shorter localization length. In the presence of extrinsic disorder, a shorter localization length (associated to dynamical disorder) makes it easier for charge carriers to be trapped by defects. In such a way, a stronger coupling to vibrations can result in an enhanced susceptibility to extrinsic defects, and possibly explain why –even though the quality of the organic crystals used is comparable for the different materials – charge carriers in TCNQ or $F_4$-TCNQ are fully trapped already at room-temperature, whereas in $F_2$-TCNQ they are not.

To conclude, besides the outstanding transport properties for $F_2$-TCNQ and the excellent reproducibility of the transport measurements for all systems considered, a key result of our work is the identification of the $F_x$-TCNQ family as a paradigm to investigate the most fundamental aspects of electronic transport in organic crystals, either experimentally or theoretically. Indeed, with the $F_x$-TCNQ family, we now have available for the first time a set of closely related molecules that crystallize in different structures, with only one of the resulting materials exhibiting very pronounced band-like transport. Given the complexity of the electronic and phononic band structure, and the energy ranges involved, understanding the behavior of these materials will require full and detailed first-principles microscopic calculations of key quantities, such as the coupling of electrons to the molecular motion, and the strength of the thermally induced fluctuations of hopping integrals, for which work is in progress. Comparing these quantities in very similar molecular materials will contribute to elucidate the key aspects in organic molecular semiconductors that determine the possibility to observe experimentally their intrinsic transport properties. It is only through these comparative studies that we can gain a truly microscopic understanding of transport in organic materials, and elucidate the reasons why different molecular crystals exhibit largely different carrier mobility values, ultimately providing viable engineering rules to optimize properties and performance.

**Supporting Information**
Supporting Information is available from the author.

**Acknowledgements**
The authors would like to thank R. Černý and L. Guinee, Laboratory of Crystallography, University of Geneva, for the crystallographic analysis of our materials, A. Ferreira for technical support, and S. Ciuchi for useful discussions. Y. K. acknowledges the financial support from the German Research Foundation (DFG) through the Research Fellowship KR 4364/1-1. M.G. acknowledges partial support by the Max Planck- EPFL Center for Molecular Nanoscience and Technology. M.G. and N.M. acknowledge support by a grant from the Swiss National Supercomputing Centre (CSCS) un- der project ID s337. Financial support from the Swiss National Science Foundation is also gratefully acknowledged.

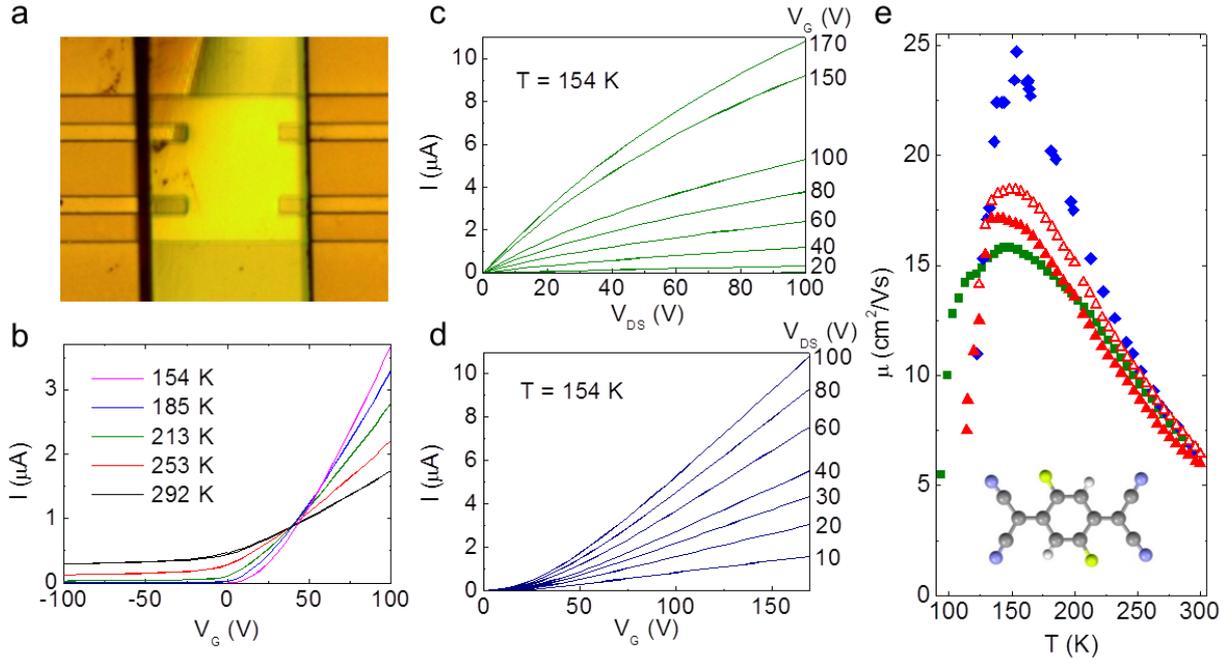

**Figure 1.** a) Optical microscope image of a vacuum-gap $F_2$-TCNQ device used for multi-terminal transport measurements. b) Transfer characteristics of a vacuum-gap $F_2$-TCNQ device measured at $V_{DS}$ = 60 V for different temperatures. c) Output and d) transfer characteristics of the same device measured at $T$ = 154 K. e) Temperature dependence of the electron mobility for three different $F_2$-TCNQ vacuum-gap devices. The mobility values were determined from the linear parts of the transfer curves at high gate voltages. Solid symbols represent the two-terminal mobility for the three devices. For comparison, open triangles show the four-terminal mobility for the same device whose two-terminal mobility is shown with full triangles (we have seen in general that in these devices the difference between two and four-terminal mobility values is negligible throughout the temperature range investigated). The inset in panel e) shows the structure of the $F_2$-TCNQ molecule.

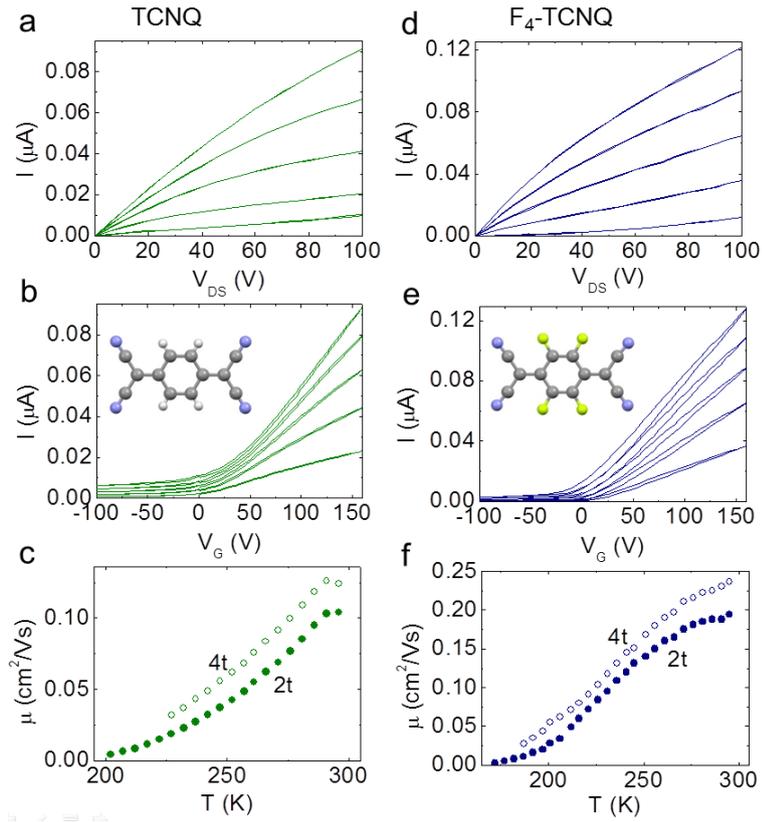

**Figure 2.** a) Room *T* output characteristics of a vacuum-gap TCNQ device ($V_G$ = 0, 40, 80, 120, 160 V). b) Room *T* transfer characteristics of the same device ($V_{DS}$ = 20, 40, 60, 80, 100 V). c) Temperature dependence of the 2-terminal and 4-terminal mobility for the same device. Panels d), e), and f) show the same measurements as panels a), b), and c), for a vacuum-gap $F_4$-TCNQ device. The insets in panels b) and e) show the structure of the TCNQ and $F_4$-TCNQ molecules, respectively.

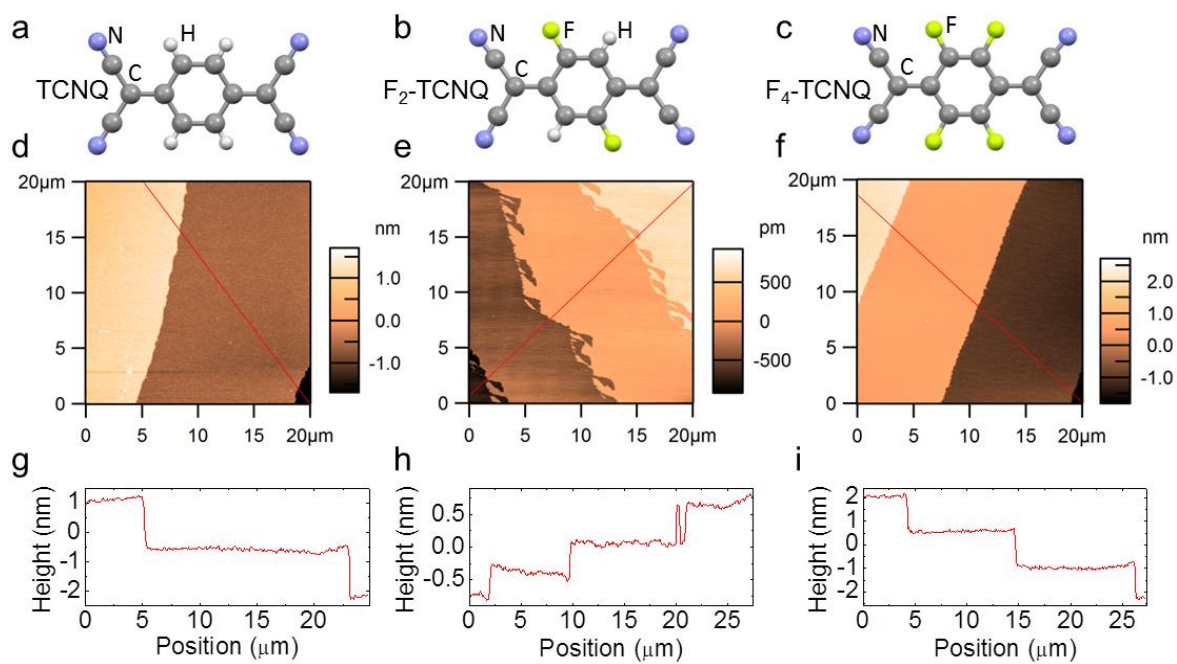

**Figure 3.** a), b), c) molecular structures of TCNQ, $F_2$-TCNQ, $F_4$-TCNQ, respectively. d), e), f) AFM images of the crystal surface for TCNQ, $F_2$-TCNQ, $F_4$-TCNQ, respectively. The height profiles along the red lines in these three figures are shown in panels g), h), and i).

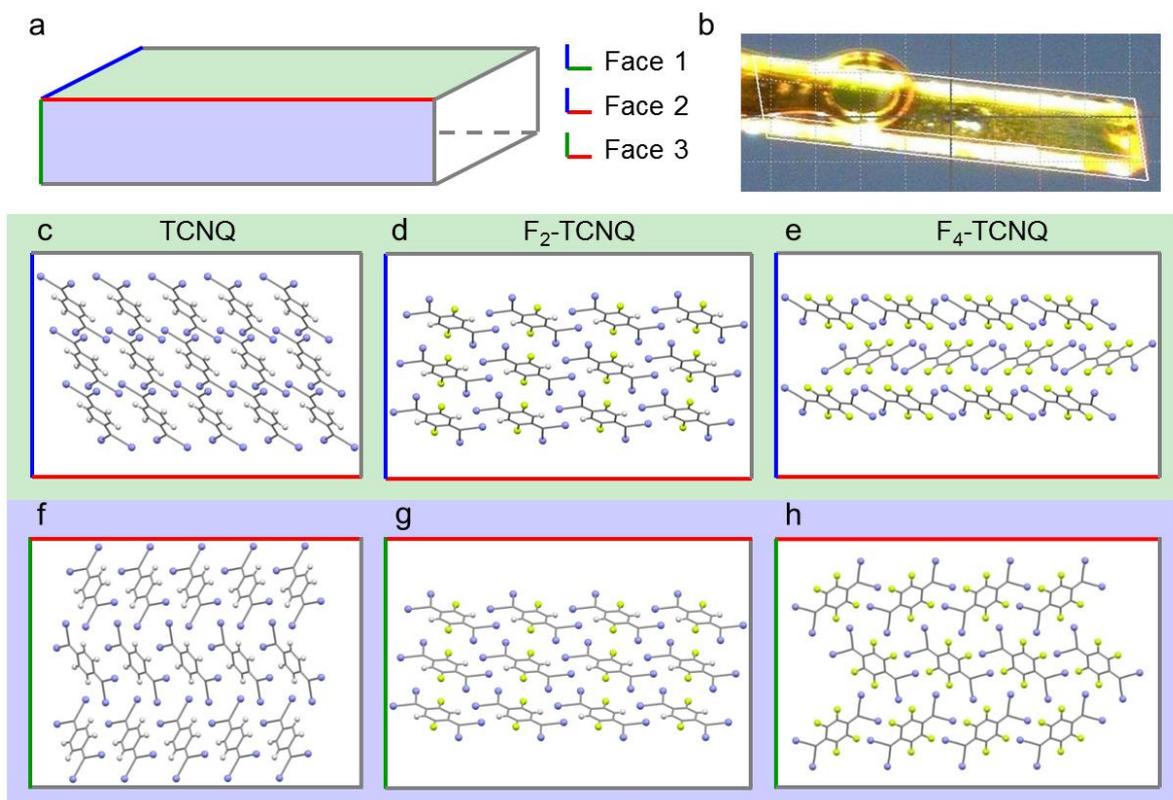

**Figure 4.** a) Schematic sketch of the crystals used in the experiments, which identifies the different relevant crystalline faces. Face 2 is the transport plane in the FET measurements. b) Optical microscope image of a crystal mounted in the X-ray diffractometer. c), d), e) packing of the molecules in the transport plane (face 2) for TCNQ, $F_2$-TCNQ, $F_4$-TCNQ crystals, respectively. f), g), h) packing of the molecules on face 3 for TCNQ, $F_2$-TCNQ, $F_4$-TCNQ crystals, respectively.

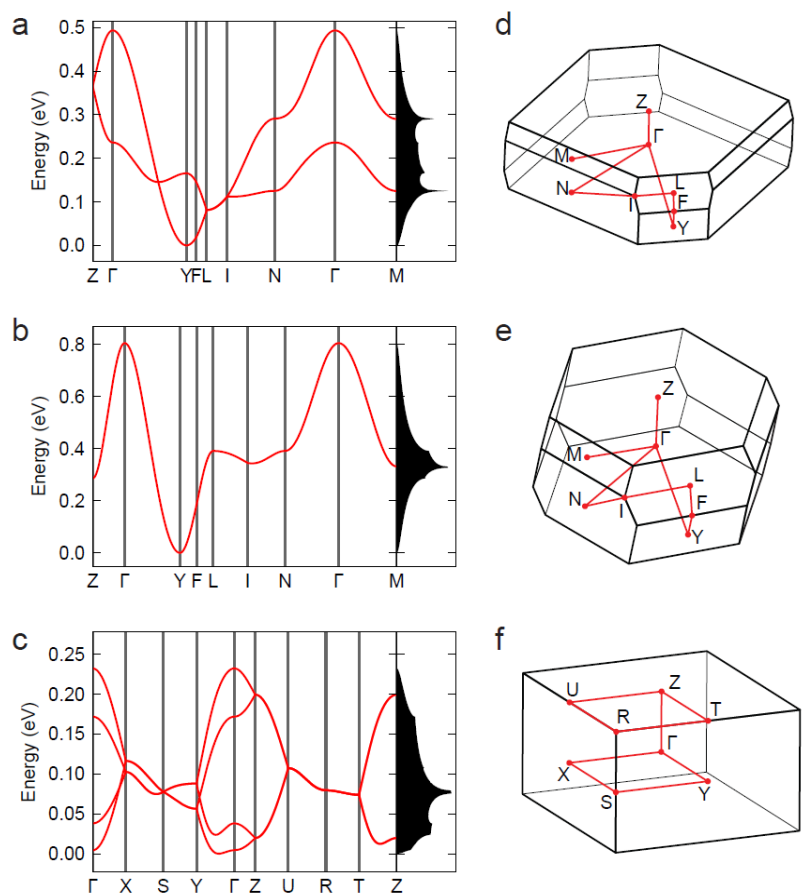

**Figure 5.** a), b), c) Band dispersion along high-symmetry directions and d), e), f) Brillouin zone of TCNQ, $F_2$-TCNQ and $F_4$-TCNQ crystals, respectively. On the right side of panels a), b), c) the density of states of the corresponding material is plotted (in arbitrary units).